\documentclass[final,5p,times,twocolumn]{elsarticle}

\newcommand{\beq}{\begin{equation}}
\newcommand{\eeq}{\end{equation}}

\newcommand{\vs}{\vskip0.15in}
\usepackage{amssymb}
\usepackage{amsbsy}  
\usepackage{hyperref} 
\hypersetup{
    colorlinks=true,
    linkcolor=blue,
    filecolor=magenta,      
    urlcolor=blue,
}
\journal{Physica C}

\begin{document}

\begin{frontmatter}
 
\title{Epilogue:  Superconducting Materials Past, Present and Future}
\author{C. W. Chu$^{a}$, P. C. Canfield$^{b}$, R. C. Dynes$^{c}$, Z. Fisk$^{d}$, B. Batlogg$^{e}$, G. Deutscher$^{f}$,  T. H. Geballe$^{g}$, Z. X. Zhao$^{h}$, R. L. Greene$^{i}$, H. Hosono$^{j}$, M. B. Maple$^{c, 1}$\fntext[fn1]{Tel.: +1 858 534 3968, $email$: mbmaple@ucsd.edu}}
\address{$^{a}$  Department of Physics and Texas Center for Superconductivity, University of Houston, Houston, TX 77004, USA \\
$^{b}$  Ames Laboratory US DOE and Department of Physics and Astronomy, Ames, Iowa 50011, USA  \\
$^{c}$Department of Physics, University of California, San Diego, 
La Jolla, CA 92093-0319 , USA\\
$^{d}$ Department of Physics, University of California, Irvine, California 92697-4574 ,USA  \\
$^{e}$ Laboratorium f. Festk\"orperphysik, ETH Zurich, Switzerland  \\
$^{f}$  Raymond and Beverly Sackler School of Physics and Astronomy, Tel Aviv University, Tel Aviv, 69978, Israel \\
 $^{g}$ Department of Applied Physics and Materials Science, Stanford University, Stanford CA
94305, USA  \\
$^{h}$ Institute of Physics, Chinese Academy of Science, Beijing 100190, China  \\
$^{i}$Department of Physics, University of Maryland, College Park, MD 20742   \\
$^{j}$  Materials and Structures Laboratory, Tokyo Institute of Technology, 4259 Nagatsuta, Midori-ku,
Yokohama 226-8503, Japan\\
}

\begin{abstract} 
Experimental contributors to the field of Superconducting Materials share their informal views on the subject.
\end{abstract}

\begin{keyword}
epilogue
\sep room temperature superconductivity
\sep perspectives



\end{keyword}

\end{frontmatter}



\section*{Introduction from the Editors}
As a closing to this {Physica C Special Issue} on Superconducting Materials,     \href{http://www.sciencedirect.com/science/article/pii/S092145341500060X} {dedicated to Ted Geballe} on the year of
his 95th birthday, we invited  selected experimental contributors to the field   to share their views on the subject, suggesting that the tone of their contribution be informal, and giving them the following guidelines:

``We suggest that you consider sharing your views on one or more of the following: (i) how the materials class(es) on which you have done experimental work fit in the \href{http://arxiv.org/abs/1504.03318}
{larger picture of superconducting materials} presented in this Special Issue, (ii) your views on any of the classes of materials and their interrelationships, (iii) what are the key unsolved questions in superconducting materials, (iv) what is the most promising route to high Tc, (v) what (if anything) should be done differently from the way it has been done to date, (vi) what you expect or hope the future will bring, e.g. whether and when we will have room temperature superconductivity, and (vii) anything else you would like to say on this topic.''

Some of the writers are authors of papers in this Special Issue,
others are not. Below are their responses. We hope these unique
perspectives from long-time major experimental contributors to
the field will be of interest to the reader and a source of
inspiration to students and researchers.

 \section*{ \href{http://www.sciencedirect.com/science/article/pii/S0921453415000878}{Paul Chu}}

 Superconductivity with impacts going far beyond condensed matter physics has lured many of the best and the brightest minds in physic sever since its discovery in 1911. It has created numerous heroes while also humbling many. Although our theoretical understanding of superconductivity in some materials, especially those with high transition temperatures, is still evolving, the experimental properties of superconducting materials have stood a much better test of time. The publication of the Physica C Special Issue on Superconducting Materials is most timely in view of the recent accelerated development of the field. It is most appropriate to dedicate this Special Issue to Ted Geballe on the year of his 95th birthday for his life-long contributions to superconductivity.
 
This Special Issue on Superconducting Materials by Physica C is unique. It is all-encompassing and authored by foremost active practitioners in the field. It summarizes all superconducting materials we know to date: from \href{http://www.sciencedirect.com/science/article/pii/S0921453415000647}{simple elements} to  {complex compounds}, and from 
 {conventional} to 
{nonconventional} superconductors.  It even includes the so-called \href{http://www.sciencedirect.com/science/article/pii/S0921453415000544}{unidentified superconducting objects} with unstable and irreproducible superconducting signals. It will serve as a great resource book for experimentalists and theorists, experts and amateurs, and veterans and beginners.      
 
From these seemingly voluminous results accumulated, similarities and differences between different families of superconductors have been drawn, various models have subsequently been proposed and further experiments have been suggested to unravel the common origin for superconductivity with high transition temperature. Unfortunately, a commonly accepted microscopic theory remains elusive, leading scientists to ponder: Are we asking the right questions? Are we on right track to getting the right information? Is it possible that high temperature superconductivity is like the common cold with more than one common cause? Is the differentiation of superconductors into different categories with an implied difference in their superconducting mechanisms proper?   The most recent report of 
detection of conventional electron-phonon mediated superconductivity 
{up to 190 K in $H_2S$ under ultrahigh pressures} appears to have raised serious challenges to our present understanding in high temperature superconductivity, if proven. At the same time, it reinforces the belief of practitioners in superconductivity that the field is vast and bright, and that more excitements are yet to come. 
 
Issac Asimov once said Òthe content of the future is what people are most insecure aboutÓ. Indeed, we may not yet know enough about the future superconductors, especially the room temperature superconductors, to ask the right questions to find them. However, we have learned from history that as long as physics does not say it will not happen, it will happen Ð superconductors at higher temperature will be discovered and new heroes will be found.
   
   \section*{ \href{http://www.sciencedirect.com/science/article/pii/S0921453415000519}{Paul Canfield}}
 After roughly 30 years in this field, I think that the best form to summarize my thoughts about new superconducting materials is the limerick.  Basically, if you cannot find humor in our fumbling about, then I fear the only alternative is insanity.
\vs
\noindent {\bf History:}
\newline
Back in the day, superconductivity was rare.
\newline
About elements and simplicity you$'$d care.

Now most material classes,

Including oxides and spin glasses,
\newline
Go super without any fanfare.
\newline

\noindent  {\bf Conventional superconductors:}
\newline
	For traditional T$_c$'s to reach high,
\newline
	Look for banding that$'$s $\sigma$ or $\pi$. 
 
		Use elements light,
 
		And bonding that$'$s tight,
\newline
	And hope critical currents wonÕ$'$t die.
\newline

\noindent  {\bf High Tc superconductors:}
\newline
	So  {copper} and 
	{iron} are done,
\newline
	Now, cobalt or nickel$'$d be fun.
 
		While the search must be agile,
 
		Look for moments quite fragile,
\newline
	And dimensions of two rather than one.
\vs

   \section*{{Bob Dynes}}
   \noindent  {\it Reflections....and Projections}
     \vs
     I grew up scientifically just after the BCS theory of superconductivity.  In that famous work and in the period following it, the quantitative verification of the role of the time-retarded, electron-phonon interaction as the mechanism responsible for the pairing of the electrons leading to the condensation into the superconducting state was a theoretical and experimental triumph. First-principles interactions described the variation of $T_c$ over a wide range of materials and experiments (tunneling, infrared, heat capacity, ultrasonic, optical light scattering, magnetic properties, electrical and thermal conductivity, NMR and others). We could describe the known phenomena very well inside the BCS framework with no real exceptions. From our knowledge, experience and confidence, we projected that no superconductor with a $T_c$ greater than 30K would be found.
     
A variety of new classes of materials were subsequently found by experimental chemists, physicists and material scientists beginning with the cuprates and followed by Fe based, B based and other compounds. They couldn$'$t have been predicted based on our (we thought thorough) understanding. We tried to force-fit these materials into our existing understanding (BCS and electron-phonon) and it just didn`t work. A more generic model of pairing which includes the electron-phonon interaction but not exclusively will come. There have been many already described but the sense of triumph after BCS and the electron-phonon interaction is still awaiting. Many of us (me included) keep searching. I look forward to that event when it happens.

 \section*{{Zachary Fisk}}

One take on superconductivity is that it shows up to solve some problem in a materialÕs electronics: good metals without problems such as Ag and Au are in general not superconductors. The cuprates and pnictides descend from compounds whose prototype compositions make ÒvalenceÓ sense and whose charged layer stacking facilitates introducing carriers. This situation is significantly different from what one finds with most metallic materials where valence is not a particularly useful concept.

In the vicinity of this ÒvalenceÓ border the materials show a variety of low temperature behaviors that suggest, loosely speaking, problems the materials are having taking care of the degrees of freedom coming from the free carriers. These carriers are interacting with a background valence bond structure which seems closer to localized chemistry than free electron physics. From this angle one sees the occurrence of superconductivity as diagnostic of a tension between bonds and bands.

	 \section*{{Bertram Batlogg}}
	    \noindent  {\it Phase-diagram-superconductors}
     \vs
While it has become convenient to classify superconductors by the 
\href{http://www.sciencedirect.com/science/article/pii/S092145341500043X}{order parameter symmetry} or the pairing mechanism as ÒconventionalÓ or Ò\href{http://www.sciencedirect.com/science/article/pii/S0921453415000660}{unconventional}Ó, an alternative view of capturing the fascination of superconductors is to realize that many are Òphase-diagram-superconductorsÓ. As an external parameter is tuned, the superconducting ground state forms next to a different electronic ground state: Mott-Hubbard insulator in layered cuprates or spin-density-wave in iron-\href{http://www.sciencedirect.com/science/article/pii/S0921453415000386}{chalcogenides}/
\href{http://www.sciencedirect.com/science/article/pii/S0921453415000477}{pnictides}. One of the earliest examples of other electronic instabilities is found in the \href{http://www.sciencedirect.com/science/article/pii/S0921453415000398}{bismuthate group} with perovskite-related structures and transition temperatures above 30K. Here, a modulation of the charge on the Bi site forms, related to the Òvalence skippingÓ tendency of Bi. A common tuning parameter is the variation of the electron count by aliovalent substitution or by intercalation, and straining the samples by external or ÒchemicalÓ pressure can also be a powerful method to map out the Òphase diagramÓ. The challenge then is to identify the underlying interactions and how the superconducting state emerges.

Whenever a new superconductor is synthesized, a first characterization may include measuring the electronic density of states at the Fermi level, the Sommerfeld $\gamma$.  Consulting the $\gamma-T_c$ ``map'' with its numerous entries of all superconductors is helpful in suggesting if the superconductor is ÒextraordinaryÓ in that the transition temperature is unusually high or low compared to others with similar density of states. Prominent outliers on the high- $T_c$ side include bismuthates, cuprates, and also \href{http://www.sciencedirect.com/science/article/pii/S0921453415000519}{$MgB_2$}, suggesting a particularly high characteristic energy of the pairing ÒglueÓ. On the other extreme one finds the 
\href{http://www.sciencedirect.com/science/article/pii/S0921453415000714}{heavy-fermion superconductors}. Interestingly, Fe-based superconductors with their particular relationship between $T_c$ and $\gamma$, are unremarkably located on the $\gamma-T_c$ ``map'', near the trend line of ,e.g., ternary 
\href{http://www.sciencedirect.com/science/article/pii/S0921453415000611}{boron-carbides} and \href{http://www.sciencedirect.com/science/article/pii/S0921453415000404}{$Nb_3Ge$}.  Thus, consulting this map might reveal unusual parameter combinations relevant for superconductivity.

It will be interesting to see if future high-$T_c$ superconductors, hopefully with application potential, will also be of the ``phase diagram superconductor'' type.

 \section*{{Guy Deutscher}}
 
 This special issue in honor of Ted Geballe on the occasion of his 95th anniversary brings back many happy memories of our never ending although geographically distant relationship. One of them is from the 1974 meeting on High Temperature Superconductivity held in Israel at the Kibbutz Kefar Gileadi, near the Lebanese border, where Ted managed to transform a threatening confrontation between Bernd Matthias and Alan Heeger into a peaceful and rewarding scientific discussion on reduced dimensionality and High $T_c$. A theme that is still with us 40 years later.
 
 One can only marvel at the rich playground that has opened up following the discovery of the 
{High $T_c$ cuprates} by Bednorz and M\"uller. The decisive step they took in their search for higher temperature superconductivity was to move from metals and alloys to nearly insulating compounds having strong correlation effects. 
 \href{http://www.sciencedirect.com/science/article/pii/S0921453415000878}{Cuprates turned out to be the right choice}. Upon doping the parent antiferromagnetic compound a phase diagram develops, running from the insulating phase to the ÒnormalÓ metallic phase, with the now famous superconducting dome in between. This phase diagram has been already studied in depth, but I feel that the two edges of the dome deserve more attention than they have so far received. 

Does the electronic structure of lightly doped, still insulating and anti-ferromagnetic samples, contain seeds of the superconducting phase? This is what the late Pierre Gilles de Gennes and I have suggested. We predicted that in this regime carriers of opposite spins could form bound pairs by contraction of Cu-O bonds around oxygen atoms, triggering the formation of charged ordered segments having a periodicity of four times the lattice constant.  It would be beautiful if higher resolution STM experiments could determine whether the predicted bond contraction actually occurs in charge ordered segments seen in lightly doped cuprates. 

At the other end of the phase diagram, are metallic-like superconducting samples really metallic in the conventional sense? Or do localized pseudo-gap states persist up to the upper edge of the superconducting dome? This question is both of academic and practical interest because localized states take away some of the superconducting condensation energy, which impacts negatively vortex pinning. Their elimination would greatly improve high temperature performance of High $T_c$ wires. It might enable superconducting magnets providing fields of the order of 10 T at liquid nitrogen temperature, not a small achievement.

Last but not least, pervading the entire phase diagram, is the question of the significance of the two distinct energy scales which I originally noted by comparing the results of Point Contact in the Sharvin limit and Tunneling experiments. The first one, the coherence energy scale $\Delta_c$, follows the same doping dependence as the critical temperature does, while the second one $\Delta_p$ increases continuously as doping is reduced. It is this second one that was first thought to be the superconducting gap a la BCS. Several beautiful experiments STM have in the mean time shown that it is highly inhomogeneous (while the DOS is homogeneous on the scale of $\Delta_c$) and there is converging agreement that it is not in fact the superconducting gap. What is still missing is a determination of the angular dependence of $\Delta_c$. Is it well defined in the anti-nodal regions of the Fermi surface, or only in the nodal ones? More detailed interference experiments between Bogoliubov-de Gennes quasi-particles, as performed by the group of Davis, might give us an answer to this question.

	 \section*{\href{http://www.sciencedirect.com/science/article/pii/S0921453415000362}{Ted Geballe}}

I thank the editors for this opportunity to share insights that I have acquired over lengthy years of research at Berkeley, Bell Labs and Stanford.  They come from a healthy mix of physics, chemistry, material science, technology and chance conversations with colleagues.

 	Why did it take so long to discover Type II superconductivity? Was it because people of my generation were taught that the high $T_c$'s [i.e. above 10K] were only found in ``hard'', i.e. brittle superconductors and were explained by Mendelsohn$'$s sponge theory? After the World War II advances in metallurgy, made in the Manhattan Project,  were declassified and ``hardness'' was identified as extrinsic, mainly due to unrecognized traces of oxygen.   MendelsohnÕ$'$s theory assumed a sponge of superconducting filaments with cross- sections less than the penetration depth, thus explaining why they had higher  $T_c$'s but
were useless for carrying superconducting currents. If the Leiden studies of PbBi alloys back in 1930 had included measurements of $J_c(H)$ the sponge theory would never have seen daylight, but amazingly it was taught and accepted for more than  25 years.  

          Advances in superconductivity come from both fundamental and applied research. The need to shield the three level microwave amplifiers [masers] to be used in the transcontinental microwave link is an example of the latter.  Rudi Kompfner asked me if 
          \href{http://www.sciencedirect.com/science/article/pii/S0921453415000404}{$Nb_3Sn$}, which we had just discovered, might be suitable.
          This open mindedness  contrasts with the negative reaction of the experts when Bernd MatthiasÕ reported the discovery of $Nb_3Sn$ in a talk at the international low temperature conference in Europe.   Rudi asked me to  test the idea.
 
  But we were fully engaged in what seemed to be a much more fundamental problem $-$ looking for a new mechanism to account for unusual superconducting behavior we had found  in transition metal compounds  that appeared to be at odds with BCS phonon-mediated pairing.  A key input to that theory, the ``isotope effect'', i.e. the square root dependence of $T_c$ upon isotopic mass,  that comes from the mass dependence of the phonon spectrum.  We noticed that it had been established only with non-transition metal elements. We obtained samples of Ru 99 and Ru 104 from Oak Ridge, and
found no dependence of $T_c$ upon mass. For the next few weeks of checking I walked around in  a state of euphoria. It ended when I walked into  Phil AndersonÕs office  and found  that our result was of fundamental importance but not because it demonstrated the need for a new mechanism, but rather  the opposite. He  pointed out that the electron-phonon interaction in narrow band superconductors is localized in space and  retarded in time, and for Ru the retardation is mass dependent and  approximately cancels the mass dependence of the phonon spectrum,  
[Phys. Rev.{\bf 125}, 1263 (1962)].  I am pleased  about our contribution even  though it didn't lead to to a new pairing mechanism it gave strong support to a deeper understanding of BCS.  The localized nature of the interaction also offered a rationale for me to understand the chemical correlation that  had puzzled me$-$why did Nb$-$containing compounds in any given family of intermetallic superconductors  always have the highest $T_c$'s?

I suggested to Rudi making $Nb_3Sn$ samples  needed for testing could best be done in the Metallurgy Dept.   The clever ``wind and then react'' method that was developed [Phys. Rev. Lett. 6, 89  (1961)] succeeded in making wires. The discovery that they carried huge currents in high fields  was totally unexpected and had major impacts in solid state and  high energy physics, in medical diagnostics, and in enabling many other technologies. AbrikosovÕs here-to-fore unappreciated theory  was ``discovered'' and type II superconductivity became main-line.

Of course many advances in superconducting science and technology come from theory.  Until graduate student Brian Josephson formulated his theory of  coherent pair tunneling across  weakly linked superconductors there was no justification for assuming that zero bias current flow was due to anything  other than shorts. That changed when phase interference effects in small  magnetic fields   were  observed by Rowell and Anderson and gave convincing proof of Josephson's theory.   The enormous impact that Josephson tunneling by paired electrons  has  had and continues to have is well known.

One day out of the blue John Bardeen phoned to ask what I thought of the Russian reports of superconductivity in the 4-6 
\href{http://www.sciencedirect.com/science/article/pii/S0921453415000489}{semiconductor PbTe} when ~1$\%$ if the Pb was replaced by Tl. My first thought was that the signals came from an undetected impurity phase but I had long learned to listen  to John and studied the Russian work carefully.  Their extensive measurements
[A. Chernik and S. N. Lykov, Sov. Phys. Solid State 23, 817 1981;   V. I. Kaidanov and Y. I. Ravich, Sov. Phys. Usp. 28, 31 (1985)] suggested new physics. 
These led to investigations in Ian FisherÕs group  that  showed thallium ions with   skipped valencies [$6S^0$ and $6S^2$] configurations are degenerate  and accompanying charge  Kondo resistance gave strong evidence for a negative U superconducting pairing mechanism [Phys Rev Lett 94 157002 2005; Phys. Rev. B 74, 134512 (2006); a history of negative U is given in arxiv.org/pdf/1406.3759]]. 

When compared with BCS superconductors that have the  same superfluid density, \href{http://www.sciencedirect.com/science/article/pii/S0921453415000362}{the $T_c'$s of PbTe(Tl) are  two orders} of magnitude greater.  Much higher  $T_c'$s are possible  if  the density of negative U pairing centers could be increased. Experiments attempting this have so far not been successful .

Synthesis and characterization capabilities have been steadily improving throughout  science and technology  [think Moore's law]. It is easy to predict that this will continue and open vast new regions of non-equilibrium phase space .  It is a matter of faith to predict that some will contain surprises that rank with  Bednorz and MuellerÕs discovery of  superconducting cuprates. I am optimistic.   I believe that some of the 
\href{http://www.sciencedirect.com/science/article/pii/S0921453415000544}{USOÕs (unidentified superconducting objects)} mentioned in earlier  chapter may  be the tip of the iceberg.  To paraphrase Vannevar Bush, our field remains a never-ending frontier.

      \section*{{Zhongxian Zhao}}
   \noindent {\it Route towards high critical temperature superconductors}
          \vs
     Prof . Ted Geballe was the first American I met in my life. It was 1975 when I was a visiting scientist in University of Cambridge. Dr. Jan Evetts introduced me to him, calling me as a native Chinese from Beijing. Ted told me that he was preparing to visit Beijing. I felt very lucky for meeting Ted because half a year later I decided to engage in exploring high critical temperature superconductors and he is the well-known authority in the field. Putting it in a Chinese way, I felt this must be a fate. We met many times over the past forty years, with a main topic in our conversations on how to search for new superconductors.

In the early days of my research, I came to form two main viewpoints on exploring high temperature superconductors based on literature search and deep thinking.  The first viewpoint is that the superconducting critical temperature is possible to exceed 40 K based on the electron-phonon interaction mechanism. The second viewpoint is that there could be other mechanisms to realize superconductivity in addition to the electron-phonon interaction (Zhongxian Zhao, Exploration of High Critical Temperature Superconductors, Physics (in Chinese) 6 (1977) 211). I still hold on these two views today. Strong electron-phonon interaction leads to lattice instability; the stronger the electron-phonon interaction, the higher the critical temperature, as long as the lattice is not undergoing a phase transition. When J. Bednorz and K.A. M\"uller reported possible high temperature superconductivity in copper-oxide compounds in 1986, I quickly realized its importance because it resonated with my viewpoints: dynamic Jahn-Teller effect can lead to lattice instability but without phase transition, thus making it possible to realize high temperature superconductivity. Now we know that the superconductivity of copper-oxides cannot be accounted for by only considering the electron-phonon interaction. The exploration of high temperature superconductivity in {metallic hydrogen}, a prediction based on the electron-phonon coupling mechanism, is promising. With an aid of some catalysis, I think it is possible to realize superconducting metallic hydrogen in some 
 \href{http://www.sciencedirect.com/science/article/pii/S0921453415000441}{hydrogen-rich compounds} under a pressure of a few hundred GPa. In general, intuition and empirical understanding can be helpful in finding new superconductors. New superconductors can be explored and discovered in materials with multiple interactions (such as 
\href{http://www.sciencedirect.com/science/article/pii/S0921453415000507}{charge order} and 
\href{http://www.sciencedirect.com/science/article/pii/S0921453415000374}{spin order} and 
\href{http://www.sciencedirect.com/science/article/pii/S0921453415000520}{tetragonal layered structures}).

The experimental data on the \href{http://www.sciencedirect.com/science/article/pii/S0921453415000477}{iron-bas}\href{http://www.sciencedirect.com/science/article/pii/S0921453415000386}{ed superconductors} are not systematic enough to extract general and key information for understanding the superconductivity mechanism. In particular, superconductors based on FeAs- and FeSe- building blocks show quite different behaviors that lead to different theoretical explanations. Such a difference may originate partly from the sample quality issue of the FeSe-based superconductors or partly from the lack of many classes of the FeSe-based superconductors present. Recent results on the phase diagram of new 1111 system $(Li_{1-x}Fe_x)OHFeSe$ (X.F. Lu et al, Nat Mater, doi:10.1038/nmat4155 (2014); X.L. Dong et al, J Am Chem Soc 137 (2015) 66) indicate that the basic characteristics of \href{http://www.sciencedirect.com/science/article/pii/S0921453415000477}{FeAs-based} and \href{http://www.sciencedirect.com/science/article/pii/S0921453415000386}{FeSe-based} superconductors could be the same. These imply the same superconductivity mechanism for the FeAs- and FeSe-based superconductors. They may even share some common features with the 
\href{http://www.sciencedirect.com/science/article/pii/S0921453415000878}{copper-oxide} 
\href{http://www.sciencedirect.com/science/article/pii/S0921453415000635}{superconductors} such as high $T_c$, tetragonal layered structure, and similar transport and magnetic properties. Such  quasi-single-layer FeSe-based (Li1-xFex)OHFeSe superconductors offer new opportunities for superconductivity research, especially with high-quality single crystal samples.

There are still plenty of rooms for exploring superconductivity in cuprates. For example, the development of new techniques to precisely tune magnetic order and fluctuations, with an aim to control both the carrier concentration and magnetic order in 
\href{http://www.sciencedirect.com/science/article/pii/S0921453415000143}{ultrathin films} of the copper-oxide parent compounds in order to achieve superconductivity, is promising in understanding the superconductivity mechanism and searching for new superconductors. Overall, there are a lot of opportunities in novel superconductor exploration based on the cuprate and iron-based superconductors.

\href{http://www.sciencedirect.com/science/article/pii/S0921453415000556}{Interface superconductivity} has been studied for many years. With the advancement of related material preparation and experimental techniques, this topic has ushered in a new opportunity. The discovery of superconductors always comes as surprises. I believe there will be many surprises to come in the future, including the discovery of room temperature superconductors. In the meantime, the complete solution of the high temperature superconductivity mechanism should come with the emergence of new solid state physics.

 \section*{{Rick Greene}}
  
I have known Ted Geballe since 1967 when I became the first postdoc in his new laboratory at Stanford University, to which he had just moved after a distinguished 15 years of research at Bell Laboratories in New Jersey. I have now enjoyed almost 50 years of interaction with Ted, both scientific and social. He has had a tremendous influence in my life--- he is one of the finest scientists and human beings that I have ever known. Little did I know how my scientific career would be changed when I started working with Ted in the summer of 1967. In fact, it was surprising to me that Ted hired me since I knew nothing about, nor had any experience with, the research that he wanted to do at Stanford on superconducting and other materials with novel electronic properties. My PhD training had been on the optical properties of magnetic insulators. So here was Ted with a completely empty lab in an old World War II building with no air conditioning, with a new postdoc who knew nothing about experimental materials research. After a few months at Stanford he may have regretted leaving the well-funded and scientifically stimulating Bell Labs, but he never let this be known to me or his newly recruited graduate students. 

The \href{http://www.sciencedirect.com/science/article/pii/S0921453415000362}{breadth of TedÕs creativity} was amazing to me! During my time as his postdoc (1967-70) we worked on many new things: granular superconductors, \href{http://www.sciencedirect.com/science/article/pii/S0921453415000507}{intercalated transition metal} 
\href{http://www.sciencedirect.com/science/article/pii/S0921453415000507}{dichalcogenide} superconductors ({reviewed by R. A. Klemm} in this issue), tungsten bronze oxide superconductors, Kondo effect materials (e.g., $SmB_6$, which has become a ÒhotÓ topic recently because of its low temperature 
\href{http://www.sciencedirect.com/science/article/pii/S0921453415000453}{topological properties}), 
\href{http://www.sciencedirect.com/science/article/pii/S0921453415000404}{A15} and 
\href{http://www.sciencedirect.com/science/article/pii/S0921453415000568}{intercalated graphite} superconductors, and a few others that I have forgotten. Some of these and many of the other superconducting materials that Ted has worked on during his long career are mentioned in this special issue of Physica C. In addition to discovering many new and interesting materials, Ted initiated the development of a new relaxation method for measuring the specific heat of small samples. Everyone in TedÕs group worked on various aspects of this idea which culminated in a famous and highly cited publication (R. Bachmann et al., Rev. Sci. Inst. 43, 205 (1972)). If you look at the author list on this paper you will see an alphabetical listing of all the people who were working for Ted during this time, and indicative of his humble nature, Ted did not put his name at the end as the senior author. Improvements of the relaxation method for specific heat measurements were developed by subsequent Geballe graduate students---e.g., \href{http://www.sciencedirect.com/science/article/pii/S092145341500060X}{Greg Stewart who has written about Ted} in this Physica C issue---and this method is now used, in modified form, in the Quantum Design PPMS systems found in many labs around the world.  So, ÒTHANKS, TEDÓ for inspiring me in my career of research on superconducting materials and in many other aspects of my life.

 So what can I say in a few paragraphs about the present status and future of superconducting materials research?  The present status is well covered by the many articles in this Physica C issue. There are crucial questions that remain unanswered about the normal state physics and the pairing mechanism in high-$T_c$ superconductors, primarily for the cuprates. The solution to the mystery of the cuprates is perhaps the most outstanding problem of condensed matter physics. Much new physics has been learned and many new or improved experimental techniques have been developed over the past 28 years of cuprate research and I expect that there will continue to be slow progress in our understanding of strongly correlated superconductors. I have my own opinions about what may be important for solving the problem of the cuprates, but I cannot justify them here so I will pass on this part of my assignment from the editors. One thing I will say is that I believe that an understanding of the 
 \href{http://www.sciencedirect.com/science/article/pii/S0921453415000635}{electron-doped cuprates} will be crucial for understanding the pairing mechanism for all of the cuprates. Recent research on the 
 \href{http://www.sciencedirect.com/science/article/pii/S0921453415000878}{hole-doped cuprates} has focused on the many interesting and perplexing properties of the ÒpseudogapÓ state. However, there appears to be no comparable ÒpseudogapÓ state in the electron-doped cuprates! This suggests to me that most of the interesting physics going on in the ÒpseudogapÓ part of the phase diagram is not directly relevant for the mechanism of superconductivity in the cuprates. Based on the research done on the electron-doped cuprates I believe that antiferromagnetic interactions of some kind (e.g. spin fluctuations or quantum critical point fluctuations) are the driving mechanism for the superconductivity in all of the cuprates. Hopefully, the theorists will soon come to some agreement on how this all works and give the experimentalists a theoretical expression for $T_c$ similar to the BCS, McMillan or Allen-Dynes formulas for the electron-phonon interaction.
 
What about the prospects for finding higher temperature superconductors?  I am quite optimistic that a room temperature superconductor will be found wihin the next 30 years, hopefully much sooner so that I can see it happen. There appears to be no theoretical reason why this is not possible. Moreover, the recent discovery of superconductivity at 190K under high pressure (Drozdoz et al., arXiv: 1412.0460), if confirmed, suggests that even the electron-phonon interaction can lead to higher-$T_c$ superconductors. Research on strongly correlated materials like cuprates, 
\href{http://www.sciencedirect.com/science/article/pii/S092145341500057X}{organics}, \href{http://www.sciencedirect.com/science/article/pii/S0921453415000581}{heavy fermions} and iron-based superconductors indicates that a magnetic pairing of some kind is necessary for higher-$T_c$. This is certainly true for these materials, but the electron-phonon interaction is back in the game for high-$T_c$.  Phonons may also be involved in the Cooper pairing for the ~100K superconductivity in single layer FeSe on STO (see I. Bozovic and C. Ahn, Nature Phys. 10, 892 (2014) for a commentary on this interfacial superconductor).
 
What is the most promising route to higher-$T_c$? I believe that we must try many different approaches. Many of the highest-$T_c$ and most interesting superconducting materials have been discovered by the ÒempiricalÓ method---educated guesswork guided by experience, intuition and hunches---with a lot of luck (serendipity). When I started searching for new superconductors with Ted Geballe in the late 1960s the 
\href{http://www.sciencedirect.com/science/article/pii/S0921453415000362}{Bernd Matthias ``rules''} were a guiding principle. Roughly stated these rules were: high symmetry is good with cubic best, high density of electronic states is good, stay away from oxygen, stay away from magnetism, stay away from insulators, and stay away from theorists. Now we have learned that these Matthias rules are mostly incorrect or too limiting. Cuprates are layered materials of lower symmetry, they are antiferromagnetic when undoped, and they are oxides! The iron-based superconductors have iron as one of their constituents and they are antiferromagnetic at low doping (or pressure). The cuprates, organics and iron-based materials not only have antiferromagnetism in their phase diagram, but they are poor metals in their normal state. 
 
So, what should be the Ònew rulesÓ for finding new superconducting materials?  I would propose:
 
1) {magnetism} is good, i.e., having a phase diagram where magnetism can be suppressed by doping or pressure;
 
2) \href{http://www.sciencedirect.com/science/article/pii/S0921453415000490}{layered structures} are good;
 
3) controlled interfaces between metals and insulators (or \href{http://www.sciencedirect.com/science/article/pii/S0921453415000489}{semiconductors}) are good;
 
4) having light elements in the structure is good, i.e., high energy optical phonons are good if mobile electrons can couple to them.
 
To reach higher $T_c$ it will be necessary to have a large J (exchange energy) in superconductors driven by a magnetic mechanism. But, lower $T_c$ new materials should not be ignored since modest structural or chemical changes or high pressure could lead to much higher $T_c$. Superconductors discovered at 
\href{http://www.sciencedirect.com/science/article/pii/S0921453415000623}{high}
 \href{http://www.sciencedirect.com/science/article/pii/S0921453415000593}{pressure} are important because it may be possible to stabilize them at atmospheric pressure via chemistry for bulk samples or by interfacial strain in a film.
 
Now what about the Matthias rule Òstay away from theoristsÓ? This was mostly true in his day, but now I would strongly recommend that experimentalists pay close attention to theorists and even enlist their help. First of all, it is helpful to have a working hypothesis for where to search for higher $T_c$ materials even if it is wrong. The discovery of high-$T_c$ superconductivity in the cuprates is an excellent example of this. The recent high-$T_c$ found in 
\href{http://www.sciencedirect.com/science/article/pii/S0921453415000441}{compressed $H_2S$ }
suggests that theoretical predictions of high-$T_c$ superconductivity can occasionally be correct, even if not exact in all the details or the predicted $T_c$ value. Second, computational methods for calculating band structure and predicting superconducting interactions are vastly improved since the days of Matthias.  In conclusion, I think that the future is bright for significant fundamental advances in superconducting materials research. 
 
    \section*{ \href{http://www.sciencedirect.com/science/article/pii/S0921453415000477}{Hideo Hosono}}
 \noindent {\it Toward Room Temperature Superconductors}
      \vs
It is a dream for human beings to realize a room temperature superconductor since the discovery of superconductivity by Heike Kammerling Onnes in 1911.  Although the fundamental theoretical framework was established in 1957 by BCS theory, there exists no theory which can quantitatively predict the critical temperature ($T_c$) even now.  Thus, exploration of high $T_c$ superconductors is like a voyage in a big ocean without a precious compass, i.e., the researchers have to determine the course believing their materials sense and/or intuition referring to what the theorists says.  In this sense, exploration of high $T_c$ superconductors is a really challenging subject in condensed matter research.  Most people would agree that this is a typical ``all-or-nothing'' research subject.
 
It is a historical fact that a material leading to a breakthrough has been discovered in most cases by chance, amidst concentrated (and often uniquely-styled) research efforts, where the researcher kept an open and flexible view of the possible routes to a new material with interesting properties. This methodology appears to be particularly true for superconductors. Here, please allow me to introduce my experience as a case study, on which I now elaborate.  In early 2009, the Japanese Government announced the launch of a new funding program, FIRST. My research proposal, ``Exploration for novel superconductors and relevant functional materials, and development of superconducting wires for industrial applications'' was fortunately accepted as one of 30 projects covering a wide range of science and technology areas. It was my expectation that novel functional materials with high potential could be found through this tough but really worthwhile challenge. I organized a research team mainly composed of Japanese solid state chemists who have much experience and a clear record of achievement in superconductors.  Since research of superconductors belongs to the domain of condensed matter physics, this team organization would be a unique feature of this project. It is my belief that excellent solid state chemists may find serendipity, even if they first fail in their hunt for the targeted new high $T_c$ materials, i.e., all or something!  Our project has examined more than 1,000 materials to explore new superconductors. The percentage of  success remained low ($~3\%$) as we estimated at first. Records of an unsuccessful trial, in a particular systematic and/or unique approach, are so informative that these should be regarded as treasures of the community.   However, no such records are 
available to the community as far as I know. Here I want to propose that we open the record of unsuccessful trials in easy accessible media such as journals, books, and data bases. We are going to publish a list of materials examined in this FIRST project in an open-access journal [H.H. Hosono et al., Sci. Tech. Adv. Mater., 2015 (in preparation)].
 
Discovery of a breakthrough superconductor has triggered the paradigm shift for new high $T_c$ superconductors.  The most representative example is the shift 
triggered by the discovery of high $T_c$ cuprates from non-magnetic intermetallic compounds with cubic symmetry to doped Mott insulators in layered transition metal oxides.  High $T_c$ cuprates were completely opposite to the working hypotheses for exploring new materials proposed by Bernd T. Matthias [W. E. Pickett, 
Physica B 296 (2001) 112].
 
   In 2010 Igor Mazin [I. I. Mazin, Nature 464 (2010) 183] proposed new working hypotheses to explore new high $T_c$ superconductors by taking discovery of high $T_c$ cuprates and Ferro pnictides into consideration;
   
(a) Layered structures are good.
   
(b) The carrier density should not be too high (compared with conventional metals)
   
(c) Transition metals of the fourth period (V, Cr, Mn, Fe, Co,Ni, and Cu) are good.
   
(d) Magnetism is essential.
   
(e) Enlist theorists, at least to compute Fermi surfaces
   
A corollary to these rules: Materials of interest are likely to be complex chemical
compounds - work closely with solid-state chemists.
 
One notes that when one sees the plot of $T_c$ vs. year that there are two modes, a continuous mode based on the conventional pairing mechanism and an abrupt mode due to a non-conventional mechanism.  What is the next example of a breakthrough superconductor?  
A striking result was posted on the online Archive on Dec. 2, 2015. The preprint reports the observation of zero-resistivity at 190K in $H_2-H_2S$ under $\sim$20GPa [A.P. Drozdov, M.I. Eremets, I.A. Troyan, arXiv:1412.0460].  If this finding comes from true superconductivity, this seems to be a straightforward approach based on the conventional BCS framework, i.e. via the formation of 
\href{http://www.sciencedirect.com/science/article/pii/S0921453415000441}{metallic hydrogen}.
 
Last, I would like to stress the fertility richness of materials we are engaging in. Exploration of novel superconductors needs a non-conventional approach
in various aspects such as the material system and synthetic processes such as high pressure and field effect.  As a result, there should be much higher probability than in other materials research to find new functionalities or to encounter new phenomena during the research.  A well-known example is the finding of high performance thermoelectric material, $NaCo_2O_4$ in the course of a comparative study of high $T_c$ cuprates with \href{http://www.sciencedirect.com/science/article/pii/S0921453415000374}{layered cobaltites}
 [I. Terasaki, Y. Sasago, K. Uchinokura, Phys. Rev. B 56 (1997) R12685(R)]. In our immediate case, excellent catalytic properties [M.Kitano et al. Nat. Chem. 4 (2012) 934] for ammonia synthesis from $N_2$ and $H_2$ gas at an ambient pressure and decomposition of $NH_3$ were found in inorganic electride C12A7:e (where C:CaO, A:$Al_2O_3$) which exhibits a bulk superconductivity at 0.2-2.5K [H. Hosono et al. Phil.Trans. Roy. Soc. A 373 (2015) 20140450]. I believe an extensive effort of exploration for new superconductors would lead to the discovery of new functionality even if a high $T_c$ material could not be targeted, because materials have huge potential and only a part of it is exposed to readily visible places (the relation between graphene and graphite is a typical example). I believe extensive exploration of new superconductors would be an all or something$-$type research 
if researchers watch for various aspects of materials.
   
   \section*
{\href{http://www.sciencedirect.com/science/article/pii/S0921453415000714}{Brian Maple}}
It is very fitting that this Special Issue on Superconducting Materials is dedicated to Ted Geballe, one of the pioneers of the field of superconducting materials.  Ted worked closely with another pioneer of this field, the late Bernd Matthias, who is known for having synthesized hundreds of new superconducting, ferromagnetic and ferroelectric materials and for his outspoken views concerning superconducting pairing mechanisms and the search for high temperature superconductors. I was associated with Matthias during most of the period he was a Professor of Physics at the University of California, San Diego, from 1960 to 1980, when he unexpectedly passed away, as a graduate student, postdoctoral research physicist and faculty colleague. Thus, it seemed appropriate to make a few remarks about Matthias and his laboratory at UCSD and a set of empirical observations about superconducting materials known as ÒMatthiasÕ rulesÓ.

Matthias had a strong personality and a charismatic manner.  Working with him wasnÕt always easy, but it was always interesting!  He taught us about the role of the periodic table of the elements in developing novel materials that serve as vehicles for studying challenging problems in condensed matter/materials physics. One of the most important things we learned from him was the excitement of research and the thrill of discovery. We met and helped entertain many of MatthiasÕ distinguished visitors; for example, John Bardeen, Ted Geballe, Ivar Giaever, Herbert Fr\"ohlich, John Hulm, and Willy Zachariesen.  Matthias traveled extensively between UCSD, Bell Laboratories and Los Alamos National Laboratory to work with his students, postdocs and collaborators, many of whom are mentioned in the Biographical Memoir written by his long time colleagues and friends, Ted Geballe and John Hulm, for the National Academy of Sciences.  In fact, a number of MatthiasÕ former students from these times have prepared articles or comments for this Special Issue, including C. W. Chu, Zachary Fisk, George Webb and myself. 
 
My Ph.D. thesis research was on the interplay between superconductivity and magnetism, which is a common thread that runs through much of my research throughout my career. My entry into this subject was inspired by the 
\href{http://www.sciencedirect.com/science/article/pii/S0921453415000532}{pioneering research of Matthias and his coworkers} who first addressed the question of whether superconductivity and magnetism could coexist in 1958.  This turned out to be a very interesting and fruitful direction of research, which led to the discovery of many new phenomena that are described in the article 
\href{http://www.sciencedirect.com/science/article/pii/S0921453415000908}{``Conventional magnetic superconductors''} by Wolowiec et al. in this Special Issue.  Subsequently, unconventional superconductivity was discovered in heavy fermion compounds, cuprates, and iron pnictides/chalcogenides.  The unconventional superconductivity often occurs in proximity to and sometimes coexists with antiferromagnetic order, where both phenomena share the same set of electrons.  Many researchers believe that the pairing of superconducting electrons in these materials is mediated by magnetic excitations. It is noteworthy that the materials with the highest $T_c$Õs, the cuprates and Fe-pnictides/chalcogenides, contain copper oxide and iron pnictide/chalcogenide layers, respectively, that can be doped with charge carriers, which suppresses the antiferromagnetic order and renders these compounds superconducting. 
 
Much has been made of MatthiasÕ ``rulesÓ'' which were based on investigations of a vast number of superconducting elements, alloys, and compounds.  These ``rules'' could be used to guide searches for new superconducting materials, particularly those with high values of $T_c$.  A ``loosely formulated'' statement of MatthiasÕ ``rules'', based on a version due to Warren Pickett, reads as follows: (1) Transition metals are better than simple metals, (2) valence electron per atom ratios of 5 and 7 are favorable, (3) high symmetry is favorable, especially cubic, (4) avoid oxygen, (5) avoid magnetism, (6) avoid insulators, and (7) avoid theorists.  While rules (1), (2) and (3) are well founded, the justification for ``rules'' (4) through (6), and, especially (7), is not so clear.  With regard to ``rule'' (5), it is noteworthy that Matthias actually proposed a magnetic pairing mechanism in 1963, based on experiments in which substitution of Fe into Ti was found to raise the $T_c$ of Ti much more rapidly than expected from his ``rules''.  (Actually, the Fe additions stabilized another nonmagnetic phase of Ti with a higher value of $T_c$.) While the interpretation of this experiment was incorrect, such a magnetic pairing mechanism as envisaged by Matthias is widely believed to be responsible for the unconventional superconductivity in heavy fermion, organic, as well as high $T_c$ cuprate and iron-based superconductors.  With respect to ÒrulesÓ (4) and (6), Matthias was actually quite interested in oxide superconductors, several of which had $T_c$Õs as high as ~12 K (e.g., $Ba(Bi,Pb)O_3$, $LiTi_2O_4$) for several years before he passed away.  It would have been interesting to see how he would have reacted to the discovery of high $T_c$ superconductivity in the cuprates in 1986.  I believe he would have been right in the middle of it!  Finally, ``rule'' (7) appears to have been added because of Matthias' public chiding of theorists about the inability of theory to predict where to find high $T_c$ superconductors.  While it is indeed true that Matthias publicly railed against theorists, he did so to emphasize the fact that theory had not been useful in predicting where to find high $T_c$ superconductors and for effect. In fact, he interacted with a number of distinguished theorists, many of whom were also his friends, such as Bardeen and Fr\"ohlich, mentioned before, as well as Phil Anderson, Al Clogston, Harry Suhl, Peter Wolff, and others. 
 
Progress in superconducting materials research has primarily been driven by the development of new and better materials. Although this field has been declared to be ÒdeadÓ at many junctures, it has proven to be  very resilient.  Time after time, new and unexpected types of superconducting materials have been discovered that ÒreenergizedÓ the field; e.g., A15 superconductors, 
\href{http://www.sciencedirect.com/science/article/pii/S0921453415000465}{magnetically ordered superconductors}, 
\href{http://www.sciencedirect.com/science/article/pii/S0921453415000416}{organic} 
\href{http://www.sciencedirect.com/science/article/pii/S0921453415000428}{superconductors}, cuprate superconductors, magnesium diboride, iron pnictide and chalchogenide superconductors.  On the basis of past experience, it seems likely that the field of superconducting materials will continue to yield new and unexpected surprises in the future.   Perhaps even room temperature superconductivity!  

\vs
\vs
\vs
In closing, the editors would like to extend their heartfelt thanks to all the authors of this Special Issue for taking time out of their busy schedules to prepare their thoughtful contributions to this endeavor under  very tight deadlines.

\end{document}